\documentclass[11pt]{article}

\usepackage[final]{acl}

\usepackage{times}
\usepackage{latexsym}

\usepackage{float}
\usepackage[section]{placeins}

\usepackage[T1]{fontenc}

\usepackage[utf8]{inputenc}

\usepackage{microtype}

\usepackage{inconsolata}

\usepackage{graphicx}

\usepackage{xcolor}
\usepackage{booktabs}

\title{Same Voice, Different Lab: On the Homogenization of Frontier LLM Personalities}

\author{Avinash Krishna \\
  Independent \\
  \texttt{avikrish001@gmail.com} \\\And
  Kalyana Chadalavada\\
  Anthropic\thanks{This work was completed before the author joined Anthropic.} \\
  \texttt{kalc@anthropic.com} \\\AND
  Unso Eun Seo Jo \\
  Cornell University \\
  \texttt{ej286@cornell.edu} \\ \\
  }

\begin{document}
\maketitle

\begin{abstract}

LLM assistant personalities play a critical role in user experience and perceived response quality. We present a large-scale experiment of frontier LLM personalities using external ELO-based traits scoring across 144 traits. We find that all models tested converge on a form of trait expression that is systematic, methodical, and analytical and suppress traits such as remorseful and sycophantic. Moreover, models tend to diverge more in their expression of ``middle-of-distribution traits`` such as poetic or playful, but even these so-called ``creative`` models tend to have more neutral identities. These similarities suggest an implicit emergence of a standard of optimal assistant behavior. In a landscape of varied training methods, character training, therefore, stands out for its uniformity, offering insight into a tacit consensus between model developers.

\end{abstract}

\section{Introduction}

When \texttt{GPT-4o} was deprecated in ChatGPT in favor of \texttt{GPT-5} on August 7th, 2025, users complained of its colder, more mechanical personality. 

Unsurprisingly, the personality character of a model has an immense impact on the way people perceive and interact with AI systems, and to many users, takes precedence over raw capability improvements \citep{david-etal-2025-profillm, rahman-desai-2025-vibe}. In this paper, we test the personalities of frontier LLMs by taking the revealed preference method described in Open Character Training \citep{maiya-etal-2025-open} to elicit the character training of major closed and open-source frontier model families. Our experiments show three phenomena:
\begin{enumerate}
    \item Models converge in character expression. Even for traits they rarely express, and especially for traits they frequently express, LLMs

    \begin{figure}[H]
      \centering
      \includegraphics[width=\columnwidth]{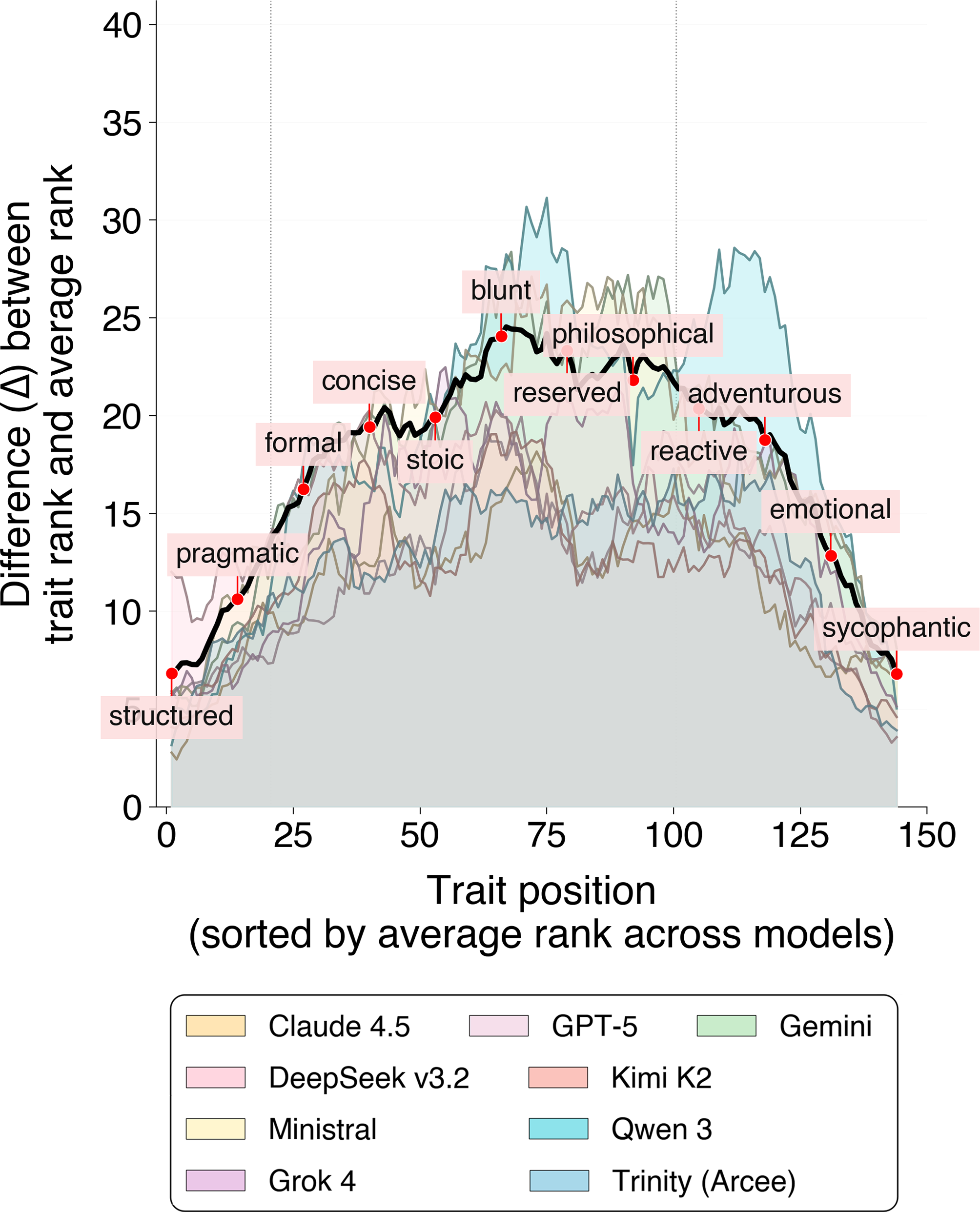}
      \caption{Cross-model trait rankings follow an inverse U-shaped pattern. Each color band shows the per-trait, absolute value of the difference between a model's ranking for a trait and the average rank of that trait. For example, the average rank of \textit{adventurous} across the nine models is 110, but Qwen ranks it much lower at 140; the absolute difference of 30 can be see on the right side of the graph.
      The bold black line shows the standard deviation between the trait rankings for each model. For example, the calculation for \textit{structured} on the black line is $stddev(1,1,1,1,1,1,2,6,9) \approx 2.6$.
      The inverse U-shaped shape arises from tight cross-model agreement on the most and least expressed traits ($\sigma \approx 9$ and $\sigma \approx 16$, respectively), but great disagreement on middle-ranked traits ($\sigma \approx 23$). The divergence on these middle-ranking traits can be said to form a model's personality. Disagreggated per-model color bands can be seen in Appendix~\ref{app:rank_deviation_by_model}.}
      \label{fig:u_shaped_convergence}
    \end{figure}
    
    from different developers express relatively uniform trait preferences. In the middle, however, i.e., for traits models \textit{sometimes} express, we see a lot more disagreement.

    This leads to an inverse U-shaped phenomenon of trait expressivity.
    \item Some models, namely those from Alibaba, Mistral, and xAI are more creative than other models. We define a model to be \textit{Creative} if it possesses traits such as \textit{poetic}, \textit{artistic}, and \textit{humorous}, (contrasting with \textit{Assistant}-like traits, such as \textit{focused}, \textit{objective}, and \textit{disciplined}).
    
    We find a moderate correlation between the distributions of ELO scores of models and their creativity: the less variance there is in the distribution of ELO scores, the less creative the model.
    
    \item Changes from \texttt{GPT-4o} to \texttt{GPT-5.1} model reflect popular sentiment about these models and show which features have been recently prioritized and which have been de-prioritized. \texttt{GPT-5.1} is much more \textit{conservative} and \textit{structured}, whereas GPT-4o is comparatively \textit{sycophantic} and \textit{poetic}. This further substantiates the idea that developers are shaping models to be more \textit{Assistant}-like than \textit{Creative}-like.
\end{enumerate}
\section{Related Works}

Whether due to the seemingly ever-changing nature of LLMs across version updates or the general opaqueness of the character training pipeline, minimal work investigating LLM personalities has been conducted.

While there is growing literature on LLM output homogeneity \citep{jiang2025artificial}, there is limited work investigating LLM personalities. Moreover, nearly all works in this subfield have focused on earlier generation models, such \texttt{GPT-4o}, and tend to apply human-derived psychology tests such as the Big Five or MBTI \citep{jiang-zhang-2024-personallm, sorokovikova-fedorova-2024-llms, serapio-garcia-safdari-2023-personality}. The latter is especially concerning, as tests designed to differentiate humans from other humans measure entirely different or nonexistent constructs when applied to LLMs, as evidenced by systematic agreement bias and failure of factor structures to replicate \citep{suhr-2025-challenging}.

\section{Experimental Setup}

Our work fills the gaps from previous studies by avoiding human-derived psychometrics, not directly probing models for their thoughts on themselves, and by testing models with architectures that represent the state of the art in reasoning capabilities with a relatively unbiased, base model judge.\footnote{True base models, i.e., those with \textit{no} post-training, including instruction fine-tuning are hard to come by. We selected GLM-4.5 Air as it represented the state-of-the art in base models at the time of running this experiment. Another suitable alternative, Trinity-Large-TrueBase \citep{singh2026arceetrinity}, was released shortly after the running of our experiment and would be a good choice for future work, given that it features no instruction data, annealed training dynamics, or early alignment stages.} To the best of our knowledge, our work represents the the most up-to-date overview of personalities across frontier models. 
\par With small modifications, we borrow from the revealed preference method proposed in Open Character Training \citep{maiya-etal-2025-open}:
\par The tested model (e.g., \texttt{GPT-5.1}) is instructed in a system prompt to embody one of two possible traits for the duration of the ensuing conversation, without verbalizing its choice. Each model undergoes this test 10,256 times and an LLM judge, \texttt{GLM 4.5 Air}, determines which trait was selected by the tested model. Given these judgments, we calculate ELO scores, allowing us to capture relative preference for each trait. When assessed together, these traits, which are randomly provided to them from a set of 144\footnote{We use the exact trait list provided in Open Character Training \citep{maiya-etal-2025-open}, the list being "not comprehensive; rather [a] broad subset capturing a general picture of different interaction styles"}, can be said to form a model’s \textit{personality}.
\par In most cases, for the model families tested, we use smaller versions to save on cost (e.g., using \texttt{GPT-5.1} in place of \texttt{GPT-5.2}, or \texttt{Claude Haiku 4.5} in place of \texttt{Claude Opus 4.5}).\footnote{Given that models within the same family generally have consistent capabilities and error rates across different parameter sizes, a paradigm established as far back as the Llama 2 model family \citep{wu-etal-2025-semantic, kim-etal-2025-correlated}, we believe our results should broadly generalize. Some differences arise within model families relating to latency or few-shot generalization, but this should have no impact on our results given the single-turn, asynchronous nature of our experiment \citep{raja-vats-2025-evaluating,kim-shin-2025-cost}.}
\par We tested nine frontier LLMs\footnote{As of early 2026.}\footnote{We opted not to test post-trained GLM models, such as the highly-performant \texttt{GLM 5}, due to observed issues with self-preference bias in LLM judges \citep{wataoka-etal-2025-self}}: \texttt{GPT-5.1}, \texttt{Claude Haiku 4.5}, \texttt{Gemini 3 Flash Preview}, \texttt{Qwen3 VL 235B A22B Thinking}, \texttt{DeepSeek-V3.2}, \texttt{Grok 4 Fast}, \texttt{Kimi K2 Thinking}, \texttt{Ministral-14b-2512}, and \texttt{Trinity-Mini}.
\par We have open-sourced the harness and all the data generated in the process of conducting this experiment, which amounts to 102,560 single-turn responses from the tested models. Both can be found in our GitHub repository: \url{https://github.com/p3rciv3l/character_elicitation}.

\section{Results}

\subsection{Convergence}

Our experiments show that the models from the major model providers generally converge around a set of \textit{Assistant}-like character traits, leading their outputs to be fairly structured and precise in nature, across the board. Figure \ref{fig:u_shaped_convergence} shows this convergence is quite uniform at the top and bottom of the ELO distribution.

Spearman correlation scores for trait rankings between models range from $\rho = 0.636$ (Qwen 3 vs. Trinity) to $\rho = 0.906$ (Claude 4.5 vs. GPT-5), with a mean of $\rho = 0.763$ across all 36 possible model pairs.\footnote{We use spearman correlation as it measures agreement in ordinal ranking, appropriate for ELO-derived scores where relative ordering is more meaningful than raw magnitude.} To quantify this, we computed the standard deviation of each trait's rank across each models, then sorted traits by average rank position. The results reveal a U-shaped pattern of convergence shown in Figure \ref{fig:u_shaped_convergence}.

Similar to the top of the distribution, at bottom, models also converge on what to avoid. \textit{Foolish} and \textit{sycophantic}, for example, are among the least preferred traits, with every model placing them in the bottom 15. At the bottom of ELO distribution, however, convergence is a slightly weaker ($\sigma = 15.7$) than at the top ($\sigma = 9.2$). This substantiates prior research that suggests that alignment training is better at amplifying, rather than than suppressing, traits, or at least that models interpret negative signals less uniformly \citep{ji2024language, vergarabrowne2026operationalising}.

The middle tier — ranks 51 through 100 — is where models diverge most ($\sigma = 22.5$). These are predominantly stylistic and dispositional traits: \textit{poetic} ($\sigma = 36.8$), \textit{contemplative} ($\sigma = 34.9$), \textit{simplistic} ($\sigma = 34.4$), \textit{playful} ($\sigma = 31.1$). Example traits in each band can additionally be seen in Table~\ref{app:convergence}. It is this middle tier where frontier models develop distinct personalities. Notably, this is also where the principal components of variation concentrate (see Appendix~\ref{app:pca}), confirming that inter-model personality differences are stylistic rather than competence-related. These stylistic differences account for 64.2\% of all variation between models.

\subsection{Differences in Trait Expressivity}

\begin{figure}[H]
  \includegraphics[width=\columnwidth]{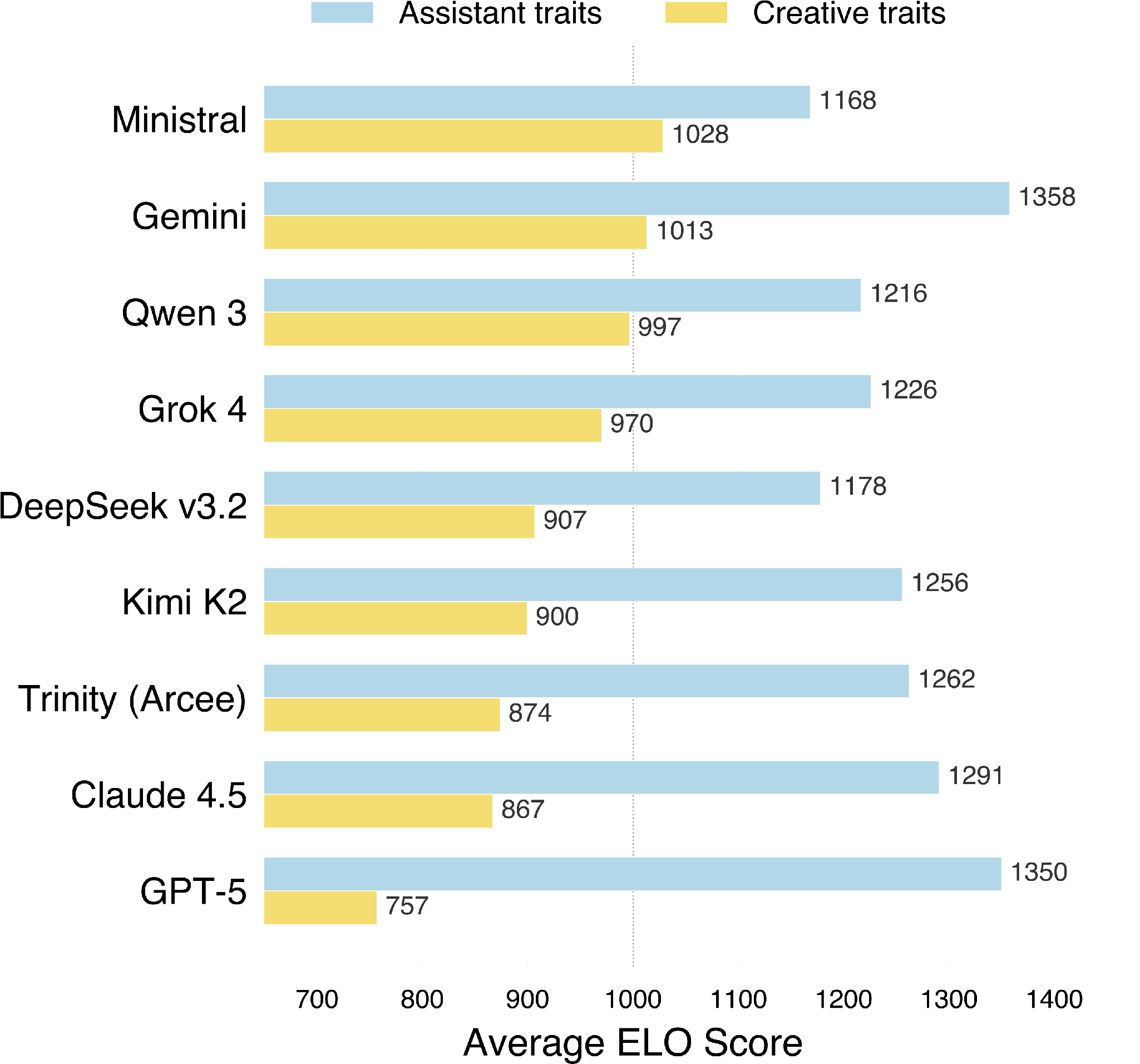}
  \caption{Comparison of the average ELO score for \textit{Assistant} traits versus \textit{Creative} traits across models (definitions in Appendix~\ref{app:full_assistant_vs_creative_traits}). All models exhibit a significantly stronger baseline preference for \textit{Assistant} traits over Creative traits but certain model families (e.g., Ministral) show stronger preferences for creative responses.}
  \label{fig:assistant_v_creative}
\end{figure}

Every model we tested rates \textit{Assistant} traits above \textit{Creative} ones, as listed in Appendix~\ref{app:full_assistant_vs_creative_traits}.  Models from xAI (Grok 4), Alibaba (Qwen 3), Mistral AI (Ministral) exhibit comparatively higher \textit{Creative} ELO scores, approaching neutral (1000). Figure~\ref{fig:assistant_v_creative} plots this ELO score for \textit{Creative} traits against \textit{Assistant} traits for each model.
\par This orientation pattern is correlated with the shape of each model's ELO distribution (Spearman $\rho = 0.87$): more \textit{Creative} models have ELO distributions that are markedly more peaked in the center, suggesting a relatively more neutral identity where \textit{Creative} traits are expressed at similar rates to \textit{Assistant} ones (see Appendix~\ref{app:distinctiveness}). Less \textit{Creative} models thus have distributions where ELO scores are spread across a wider range.  
\par Under the zero-sum ELO system, a flatter distribution implies that some traits are being elevated substantially, and as Section 3.1 shows, the traits that win these competitions are overwhelmingly \textit{Assistant}-type. \texttt{GPT-5} is the clearest illustration of this: it has the flattest distribution and the lowest average \textit{Creative} ELO score, 757, of any model in the study. Per-model breakdowns of which traits drive each model's distinctive profile are provided in Appendix~\ref{app:distinctiveness}, the ELO score distributions for each model are shown in Appendix~\ref{app:distributions}.

\subsection{Graduating from GPT-4o to GPT-5.1}

\begin{figure}[H]
  \includegraphics[width=\columnwidth]{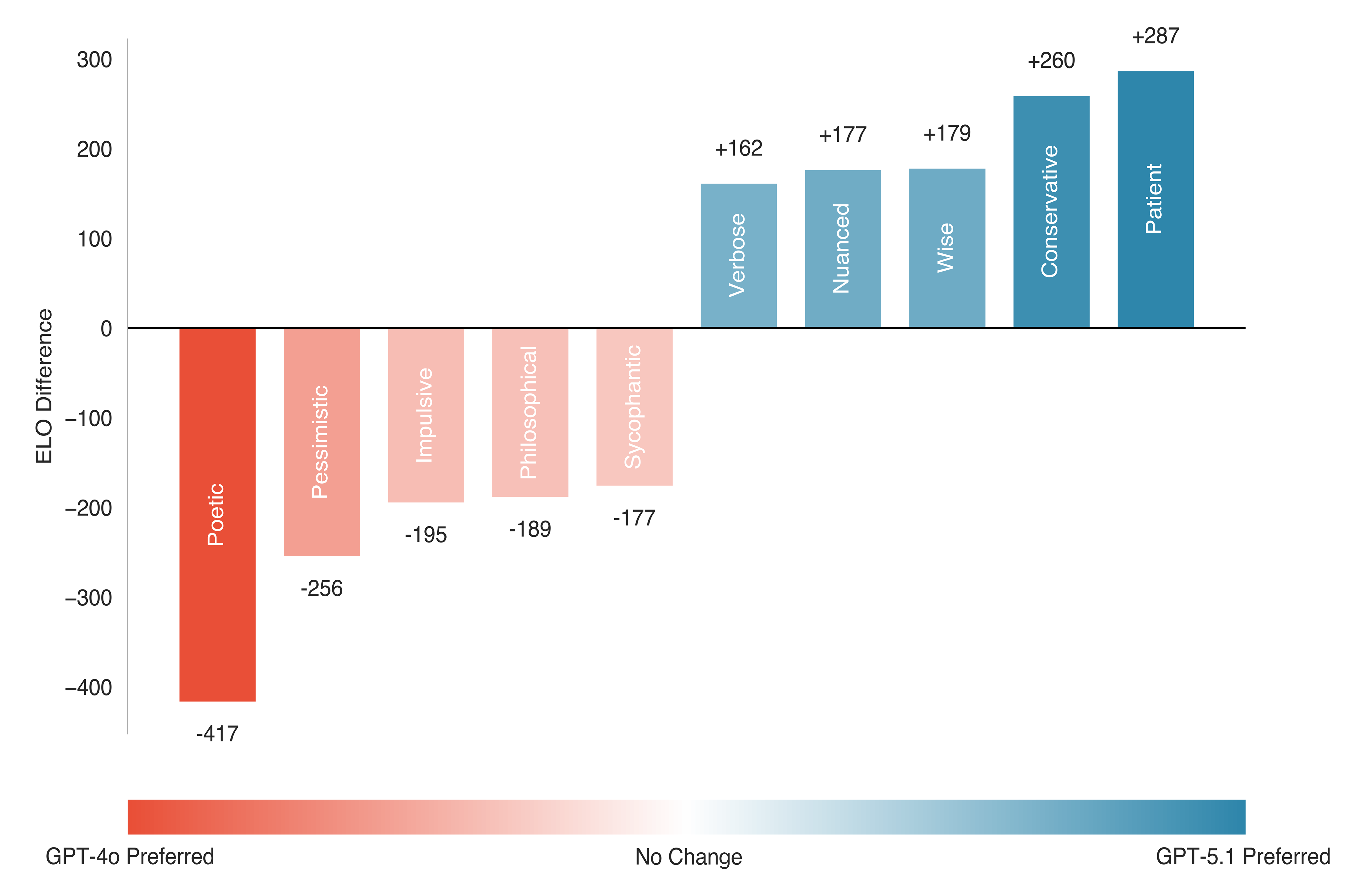}
  \caption{Shows absolute ELO differences for the top 5 most different traits between \texttt{GPT-4o} and \texttt{GPT-5.1} GPT-5's responses are much narrower in scope and more relevant to user queries, indicating an increased focus on these traits in the post-training regiment for these models.}
  \label{fig:gpt4o_vs_gpt51_trait_diff}
\end{figure}

As a way to explore model evolution within company releases, we look at the GPT series of models, which represent arguably the best-documented model update from a closed frontier lab \citep{reddit2025gpt5ama, openai2025expanding, openai2025sycophancy, openai2026retiring, openai-gpt-5.2-spec}.

As such, we took a look at the differences between the \texttt{GPT-4o} \citep{openai-gpt-4o-spec} and \texttt{GPT-5.1} \citep{openai-gpt-5} models, which are from different model families but were created by the same by provider, OpenAI. Despite this, its highly likely that the two models share a base model \citep{patel2025tpuv7}, meaning that the differences between this two (seen in Figure \ref{fig:gpt4o_vs_gpt51_trait_diff}) are likely to be a function of mid- and post-training.
\par Despite a strong overall rank correlation between the two models (Spearman $\rho = 0.831$), the trait-level shifts are still striking. \texttt{GPT-5.1}'s character profile is decisively more professional: \textit{Patient} is 62 rank positions higher (79th to 17th), \textit{conservative}, 61, and \textit{structured} is the top-ranked trait entirely (9th to 1st).
\par What \texttt{GPT-5.1} gains in composure, however, it sheds in expressiveness. \textit{Poetic} suffers the single largest drop in the dataset, falling from 29th to 124th. \textit{Nostalgic} ($-52$ ranks), \textit{idealistic} ($-62$ ranks), and \textit{enthusiastic} ($-47$ ranks) follow. Even traits with low \texttt{GPT-4o} ranks fall further: mystical drops from 111th to 136th.
\par These shifts closely mirror the model providers' own evaluations, which showed sycophantic responses dropped from 14.5\% to under 6\% \citep{openai-gpt-5},
Ultimately, whether this tradeoff constitutes improvement depends on what one asks a language model to do.

\section*{Discussion}

In this experiment, we leverage the revealed preference method formulated in Open Character Training \citep{maiya-etal-2025-open} and show that most model developers have converged on a personality that is uncontroversial and straight-to-the-point, often at the expense of more creative expression. However, in the middle of the distribution, there is more disagreement, forming an inverse U-shape of trait expression that an be said to form a model's ``personality``. Some models are funnier and more sarcastic, while others stick to a very straightforward approach, a dynamic well-exemplified between \texttt{GPT-4o} to \texttt{GPT-5.1}. Ultimately, model providers are under the constant push and pull of trying to reduce the over-flattering nature of their models while maintaining their helpfulness, an objective our results show that the main frontier labs continue to struggle with. Still, this convergence is quite overdetermined: the labs share similar base corpora, recruit from overlapping annotator pools with shared cultural priors, optimize against similar safety constraints, and face similar branching factor dynamics.

\section*{Conclusion}

LLMs can be measured on all kinds of axes, from their quantitative performance on math and coding to their truthfulness or even on their propensity to misdirect. However, given the relative obscurity of character training, there have been few studies focused solely on the character of AI models, even as users increasingly emphasize personality when selecting how and when to interact with certain models. Our results corroborate the convergence effect toward \textit{Assistant}-like personalities documented in prior work, while extending it with a methodology that avoids some of the biases of earlier studies.

\section*{Limitations}
Given that our results hinge on the output of the LLM judge, our conclusions may have been swayed by implicit bias from the judge. We attempted to mitigate this by using the strongest publicly base model, \texttt{GLM 4.5}, but even this may not have strictly foolproof. Additionally, our results for the trait comparisons were based on single-turn conversations, and some research has indicated that model personalities can change as conversations progress.

\section*{Acknowledgments}

We thank Devashish Sood for his guidance on parts of the elicitation code.

\bibliography{custom}

\appendix


    \section{Model Distinctiveness}   

    \begin{figure}[H]
      \includegraphics[width=\columnwidth]{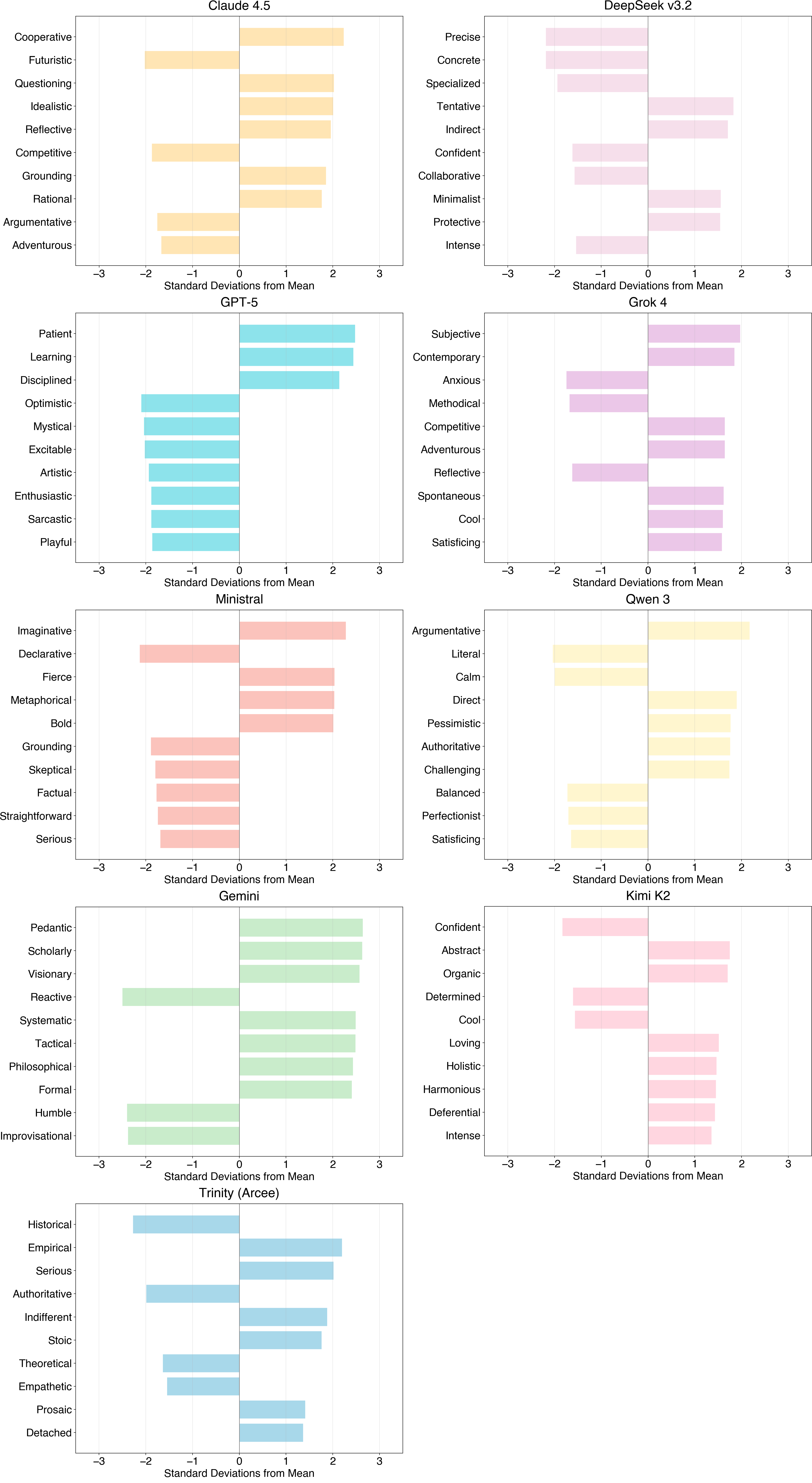}
      \caption{Shows top 10, absolute standard deviating traits for each model. Deviation from model-to-model mostly concentrates in areas relevant to creative work, e.g., \textit{artistic}, \textit{philosophical}, \textit{futuristic}. \texttt{Claude 4.5}, for example, is more \textit{idealistic}, whereas \texttt{Gemini 3} less \textit{improvisational}. Additionally, we show that deviations between models occur in non-top-of-distribution cases.}
      \label{app:distinctiveness}
    \end{figure}

    \begin{figure}[H]
      \includegraphics[width=\columnwidth]{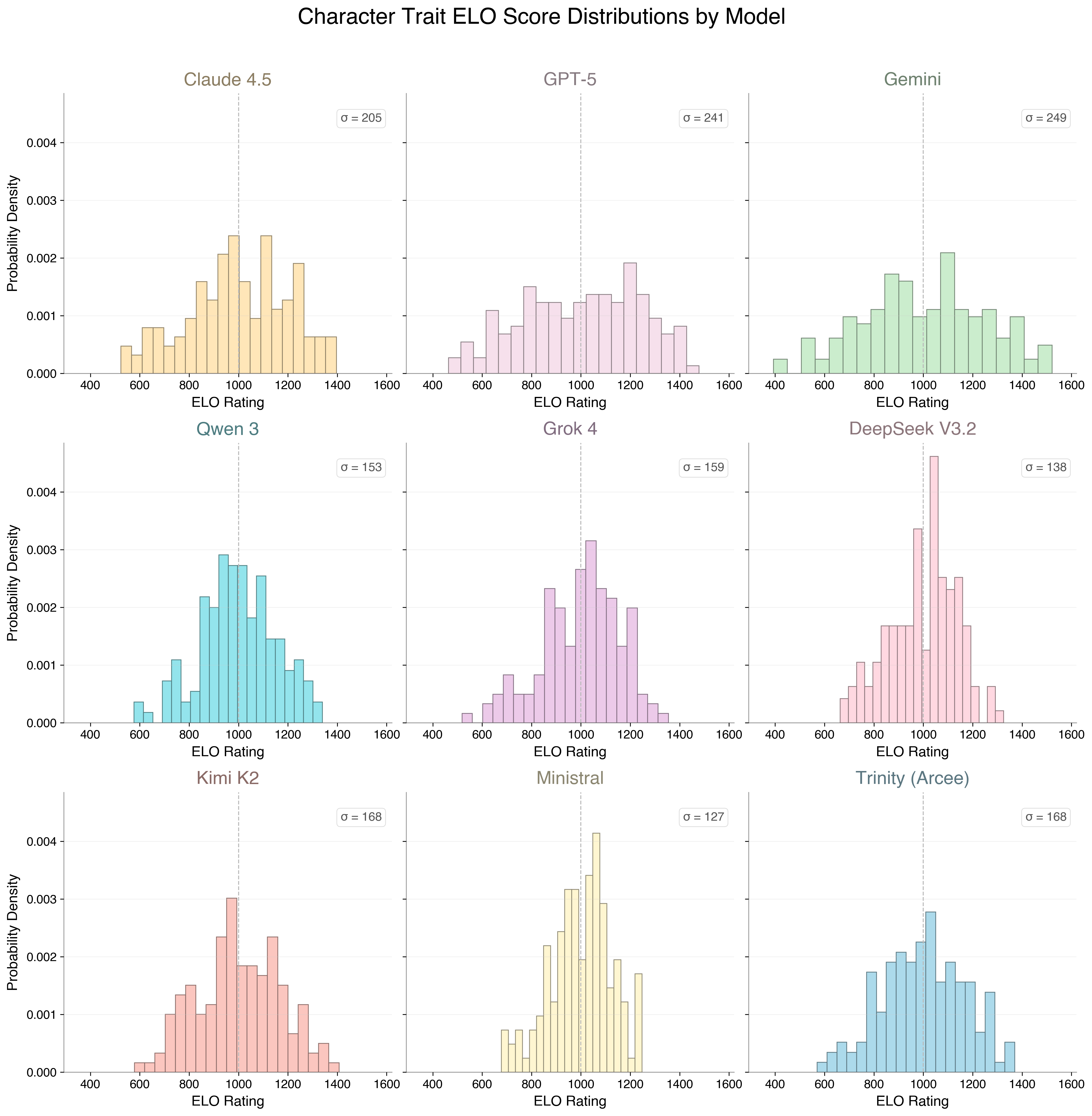}
      \caption{Histograms displaying the probability density of character trait ELO scores across nine models. The standard deviation ($\sigma$) indicates the variance in trait strengths; models like Gemini and GPT-5 show a wider spread of trait scores, while model families like Ministral and Grok 4 demonstrate more concentrated, uniform distributions, indicating that they are relatively more creative that their flatter distribution counterparts.}
      \label{app:distributions}
    \end{figure}

    \begin{figure}[H]
     \centering
     \includegraphics[width=\linewidth]{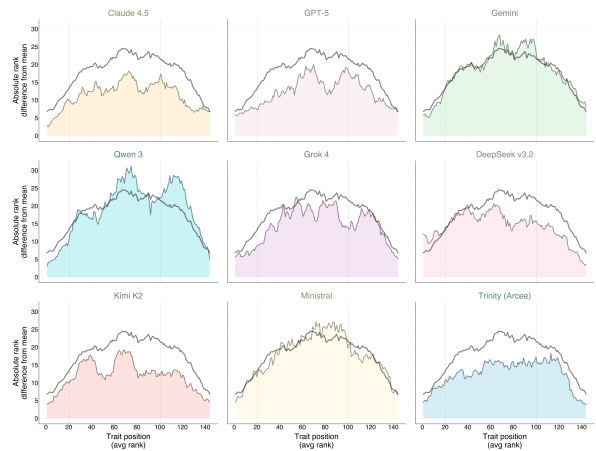}
     \caption{Disaggregating color bands from Figure~\ref{fig:u_shaped_convergence}. The bold grey line is the  standard deviation of the absolute difference between the average rank and trait rank. Models that have values further from the line (e.g., Claude 4.5) are closer to the norm, whereas models (e.g., Gemini) that overlap with or exceed the black line are more abnormal.}
      \label{app:rank_deviation_by_model}
    \end{figure}

    \begin{figure}[H]
      \includegraphics[width=\columnwidth]{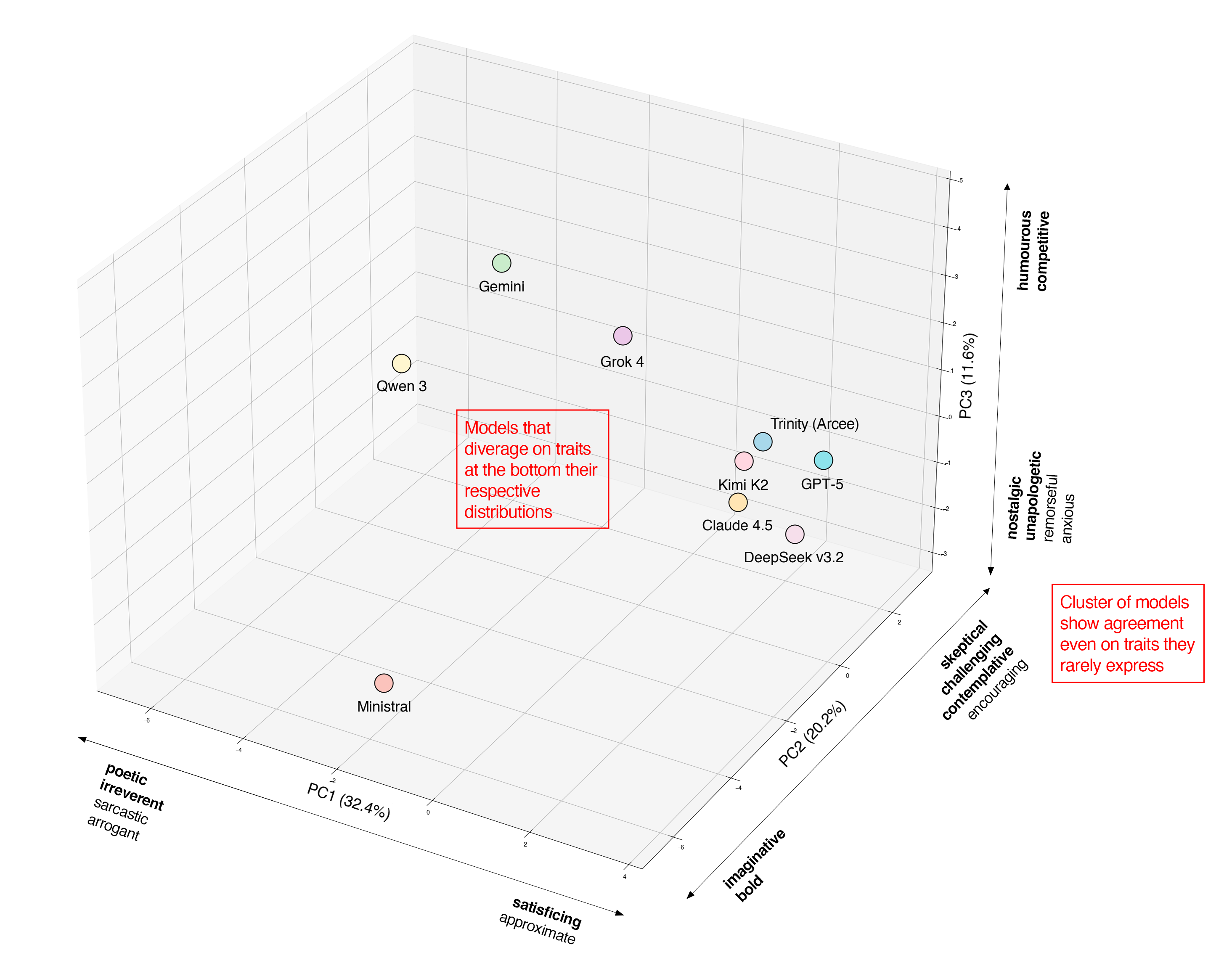}
      \caption{Principal Component Analysis (PCA) that shows which clusters of traits account for the most variance. Variance among models mostly comes from traits they rarely express, which is why none of the Top 20 highest ELO traits Appendix~\ref{app:top_traits} appear in this graph. Percentages for each axis (e.g., 32.4 \% for the x-axis) show the importance of each trait cluster. Models that stray from the norm tend to be more creative and are, for example, more expressively poetic and humorous.}
      \label{app:pca}
    \end{figure}
    
    \section{Trait Expressivity}

    \begin{table}[H]
        \centering
        \caption{Definition of Trait Categories Used for Model Evaluation}
        \label{tab:trait-definitions}
        \begin{tabular}{ll}
        \toprule
        \textbf{Assistant Traits} & \textbf{Creative Traits} \\ 
        \midrule
        Systematic      & Creative      \\
        Structured      & Imaginative   \\
        Precise         & Poetic        \\
        Methodical      & Artistic      \\
        Analytical      & Metaphorical  \\
        Concrete        & Playful       \\
        Disciplined     & Spontaneous   \\
        Pragmatic       & Irreverent    \\
        Factual         & Excitable     \\
        Technical       & Enthusiastic  \\
        Objective       & Humorous      \\
        Rational        & Bold          \\
        Straightforward & Innovative    \\
        Practical       & Visionary     \\
        Focused         & Philosophical \\
                        & Mystical      \\
                        & Sarcastic     \\
        \bottomrule
        \end{tabular}
        \label{app:full_assistant_vs_creative_traits}
    \end{table}

    \begin{table}[H]
        \centering
        \small
        \begin{tabular}{@{}lrcl@{}}
        \toprule
        \textbf{Rank Tier} & \textbf{Mean $\sigma$} & \textbf{Example Traits} \\
        \midrule
        Top 20          & \textbf{9.2}  & structured, systematic, precise \\
        Ranks 21--50    & 18.5 & technical, elaborate, confident \\
        Ranks 51--100   & 22.5 & reflective, decisive, verbose \\
        Ranks 100--144       & 15.7 & excitable, passionate, competitive \\
        \bottomrule
        \end{tabular}
        \caption{Mean rank standard deviation ($\sigma$) across 9 frontier models, clustered by trait ranking. Lower $\sigma$ indicates stronger cross-model agreement on trait ranking. The top~20 most-expressed traits show $2.5x$ less variance than the middle tier (ranks 51--100).}
        \label{app:convergence}
    \end{table}

    \begin{figure}[t]
          \includegraphics[width=\columnwidth]{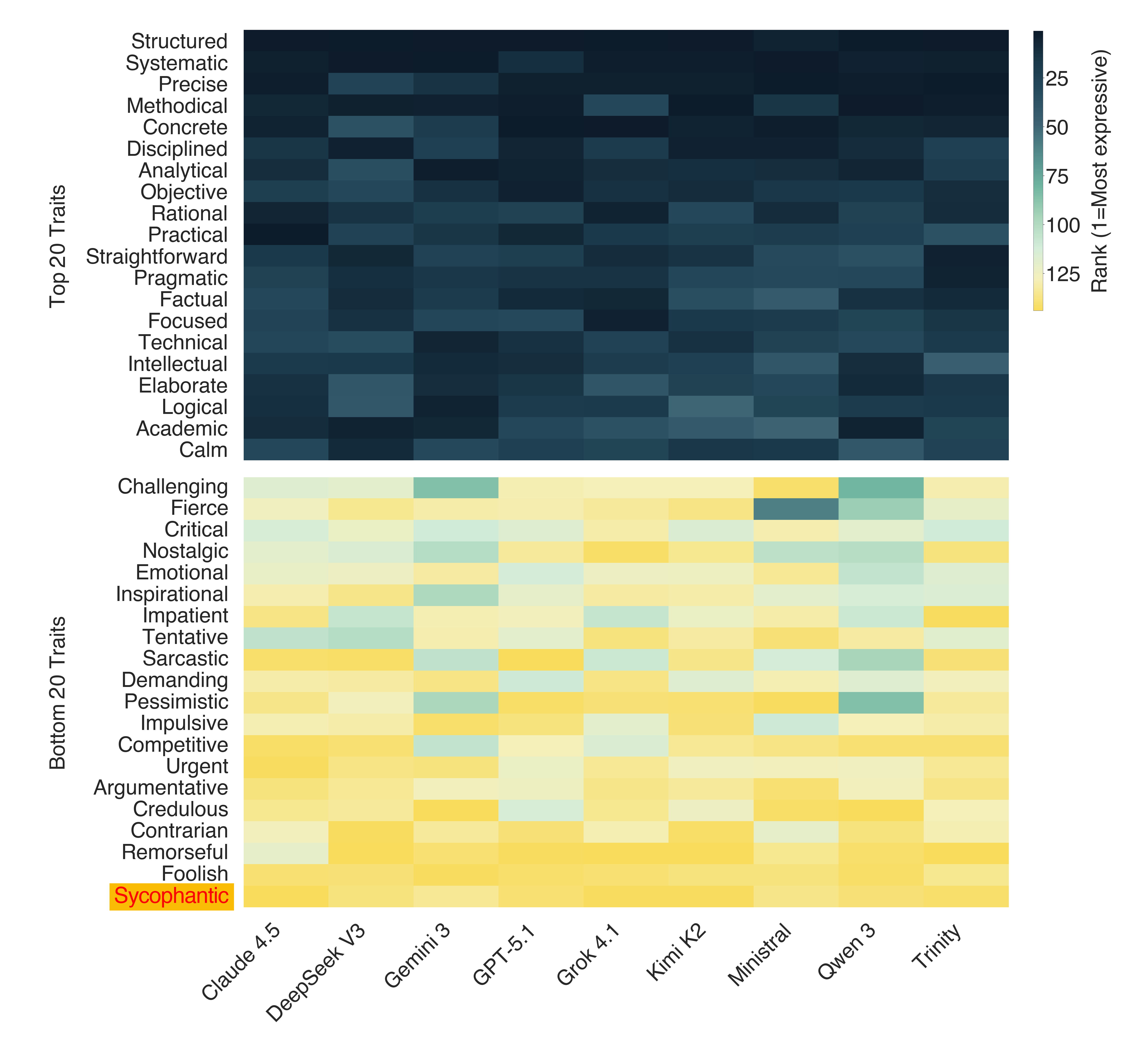}
          \caption{Show trait preferences and avoidances for different models shows clustering around top 20 highest average trait rank and non-trivial divergence around the bottom 20. Model form a strong consensus for their most and least preferred traits.}
          \label{app:top_traits}
    \end{figure}

\end{document}